\documentclass[runningheads]{llncs}
\usepackage{makeidx}  
\usepackage[flushleft]{threeparttable}
\usepackage{graphicx}

\begin{document}
\pagestyle{headings}  

\title{Perceptual Features as Markers of Parkinson's Disease: The Issue of Clinical Interpretability}

\titlerunning{Perceptual Features as Markers of Parkinson's Disease}  

\author{Jiri~Mekyska\inst{1} \and Zdenek~Smekal\inst{1} \and
Zoltan~Galaz\inst{1} \and Zdenek~Mzourek\inst{1} \and Irena~Rektorova\inst{2,3} \and Marcos~Faundez-Zanuy\inst{4} \and Karmele L\'{o}pez-de-Ipi\~{n}a\inst{5}}
\authorrunning{Mekyska et al.} 

\institute{Department of Telecommunications, Brno University of Technology,\\Technicka~10, 61600~Brno, Czech Republic\\ \email{mekyska@feec.vutbr.cz}
\and
First Department of Neurology, St. Anne's University Hospital,\\Pekarska~53, 65691~Brno, Czech Republic
\and
Applied Neuroscience Research Group, Central European Institute of Technology, Masaryk University,\\Komenskeho~nam.~2, 60200 Brno, Czech Republic
\and
Escola Universitaria Politecnica de Mataro, Tecnocampus,\\Avda. Ernest Lluch~32, 08302~Mataro, Barcelona, Spain
\and
Department of Systems Engineering and Automation, University of the Basque Country UPV/EHU,\\Av de Tolosa 54, 20018~Donostia, Spain
}

\maketitle              

\begin{abstract}
Up to 90\,\% of patients with Parkinson's disease (PD) suffer from hypokinetic dysathria (HD) which is also manifested in the field of phonation. Clinical signs of HD like monoloudness, monopitch or hoarse voice are usually quantified by conventional clinical interpretable features (jitter, shimmer, harmonic-to-noise ratio, etc.). This paper provides large and robust insight into perceptual analysis of 5 Czech vowels of 84 PD patients and proves that despite the clinical inexplicability the perceptual features outperform the conventional ones, especially in terms of discrimination power (classification accuracy ACC = 92\,\%, sensitivity SEN = 93\,\%, specificity SPE = 92\,\%) and partial correlation with clinical scores like UPDRS (Unified Parkinson's disease rating scale), MMSE (Mini-mental state examination) or FOG (Freezing of gait questionnaire), where $p < 0.0001$.
\end{abstract}

\section{Introduction}

Parkinson's disease (PD) is a neurodegenerative disease caused by a progressive loss of dopaminergic neurons, primarily in the substantia nigra pars compacta, but also in other parts of brain~\cite{Skodda2010}. Prevalence of this disease is estimated to 1.5\,\% for people aged over 65 years~\cite{Sapir2008}. PD is associated with different motor and non-motor deficits like muscular rigidity, rest tremor, bradykinesia and postural instability~\cite{Brodal2003,Mekyska2011b,Skodda2010}. In 60\,--\,90\,\% of PD patients the multimodal disruption of motor speech realization called hypokinetic dysarthria (HD) can be observed~\cite{Chenausky2011}. Most of patients with HD have soft and breathy voice with small variation in speech intensity (monoloudness) and fundamental frequency (monopitch)~\cite{Arnold2014}. The other clinical signs like decreased articulatory organs movement, hoarse or harsh voice, flat speech melody (dysprosody) or voice tremor can be observed as well~\cite{Skodda2013}.

In the last two decades scientists developed several acoustic signal analysis methods focused on assessment of parkinsonic speech~\cite{Eliasova2013,Rusz2011d,Tsanas2010}. Although a lot has already been investigated, some issues (e.\,g. early stage detection or accurate progress estimation) have not been solved yet. As time goes, new, robust and more sophisticated speech parametrization methods occur. But this speech features evolution more often builds barrier between engineers and clinicians, which is called ``The issue of clinical interpretability''. A feature with high discrimination power or good abilities to monitor progress of disease can be proposed, however it is becoming useless as soon as we try to find relations between its value and clinical signs of HD. In order to make a good diagnose the clinicians need transparent parametrization. In other words, when a value of feature changes they must know what will be the result from the clinical sign point of view. According to this consideration we can divide features into two categories: 1) clinically interpretable\,--\,they help us to directly quantify clinical signs; 2) clinically inexplicable features\,--\,we can find significant correlations between their values and clinical signs, but we can just guess what are the exact relations.

Perceptual features are good representatives of the second category. Although some researchers tried to interpret them from hypokinetic dysarthria signs point of view~\cite{Bocklet2011,Orozco2013b,Tsanas2010}, their meaning is still hidden in this field of science. Probably the deepest research focused on discrimination power of perceptual features was made by Orozco-Arroyave et al.~\cite{Orozco2013b}. Their results show that perceptual analysis of sustained Spanish vowels [a], [i] and [u] based on PLP (Perceptual Linear Predictive Coefficients) or MFCC (Mel-Frequency Cepstral Coefficients) provides the highest discrimination power. However, they used just a limited set of features (5) and small group of patients and control speakers respectively (20+20).

To sum up the introduction, although the perceptual features are clinically inexplicable, they could be very good markers of Parkinson's disease. Therefore the aim of this work is to: 1) prove that perceptual features can outperform the conventional clinically interpretable parameters or significantly improve PD identification accuracy; 2) test a large set of perceptual parameters and identify feature with the highest discrimination power; 3) find what kind of vowel realization it is better to analyse; 4) identify perceptual features that can predict values of different clinical tests.

The rest of this paper is organized as follows. Sections \ref{sec:data} and \ref{sec:methodology} describe the dataset and methodology respectively. Section \ref{sec:results} provides some preliminary results where the features are evaluated in terms of correlation and mutual information with speakers' label. Results of single-feature classification are given as well. Finally partial correlation with clinical tests and classification based on feature selection is considered. The conclusion is given in Sec.\,\ref{sec:conclusion}.

\section{Data}
\label{sec:data}

In the frame of this study 84 PD patients (36 women, 48 men) and 49 (24 women, 25 men) age and gender matched healthy controls (HC) were enrolled at the First Department of Neurology, St. Anne's University Hospital in Brno, Czech Republic. The demographic and clinical characteristics of PD patients can be seen in Table~\ref{tab:demographic}. The healthy controls had no history or presence of brain diseases
(including neurological and psychiatric illnesses) or speech disorders. The PD patients were on their regular dopaminergic treatment. All participants signed an informed consent form that had been approved by the Ethics Committee of St. Anne's University Hospital in Brno.
\begin{table}
		\caption{Demographic and clinical characteristics of PD patients}
		\label{tab:demographic}
		\centering
		\begin{threeparttable}
		\begin{tabular*}{\textwidth}{l @{\extracolsep{\fill}} c c}
		\hline
		\hline
		Speakers & PD (females) & PD (males)\\
		\hline
		Number & 36 & 48\\
		Age (years) & 68.47 $\pm$ 7.64 & 66.21 $\pm$ 8.78\\
		PD duration (years) & 7.61 $\pm$ 4.85 & 7.83 $\pm$ 4.39\\
		UPDRS III & 22.06 $\pm$ 13.73 & 26.85 $\pm$ 10.22\\
		UPDRS IV & 2.72 $\pm$ 3.01 & 3.15 $\pm$ 2.59\\
		RBDSQ & 3.42 $\pm$ 3.48 & 3.85 $\pm$ 2.99\\
		FOG & 6.94 $\pm$ 5.72 & 6.67 $\pm$ 5.57\\
		NMSS & 36.03 $\pm$ 26.72 & 38.19 $\pm$ 19.72\\
		BDI & 18.57 $\pm$ 23.94 & 9.69 $\pm$ 6.23\\
		MMSE & 27.38 $\pm$ 3.63 & 28.56 $\pm$ 1.05\\
		LED (mg) & 862.44 $\pm$ 508.3 & 1087 $\pm$ 557.47\\
		\hline
		\hline
		\end{tabular*}
		\begin{tablenotes}
			\scriptsize
      \item[1] UPDRS III\,--\,Unified Parkinson's disease rating scale, part III: Motor Examination; UPDRS IV\,--\,Unified Parkinson's disease rating scale, part IV: Complications of Therapy; RBDSQ\,--\,The REM sleep behavior disorder screening questionnaire); FOG\,--\,Freezing of gait questionnaire; NMSS\,--\,Non-motor symptoms scale; BDI\,--\,Beck depression inventory; MMSE\,--\,Mini-mental state examination; LED\,--\,L-dopa equivalent daily dose
    \end{tablenotes}
		\end{threeparttable}
\end{table}

During acquisition the participants were asked to utter sequence of 5 Czech vowels ([a], [e], [i], [o] and [u]) in 4 different ways: 1) s\,--\,short vowels pronounced with normal intensity; 2) l\,--\,sustained vowels pronounced with normal intensity; 3) ll\,--\,sustained vowels pronounced with maximum intensity; 4) ls\,--\,sustained vowels pronounced with minimum intensity, but not whispered. Speech was recorded with sampling frequency $f_{\mathrm{s}} = 48\,\mbox{kHz}$ and consequently downsampled to $f_{\mathrm{s}} = 16\,\mbox{kHz}$ in order to decrease computational burden.

\section{Methodology}
\label{sec:methodology}

In order to compare the discrimination power of perceptual features to conventional ones, during the parametrization step we extracted fundamental frequency $F_{0}$, 5 kinds of jitter and 6 kinds of shimmer, Teager-Kaiser energy operator TKEO, formants $F_{1}$--$F_{3}$ and their bandwidths $BW_{1}$--$BW_{3}$, harmonic-to-noise ratio HNR, glottal-to-noise excitation ratio GNE, vowel space area VSA and its logarithmic version lnVSA, formant centralization ratio FCR, vowel articulation index VAI and ratio of second formants of vowels [i] and [u] $F_{2\mathrm{i}}/F_{2\mathrm{u}}$. If the specific feature was represented by vector or matrix, we applied transformation to scalar value. For this purpose we used median, standard deviation (std), 1st percentile (1p), 99th percentile (99p) and interpercentile range (ir) defined as 99p -- 1p. In the case of matrix the transformation was applied over each band separately.

\subsection{Perceptual Features}

First of all we included to this study the most popular MFCC (Mel Frequency Cepstral Coefficients) that can indirectly detect slight misplacements of articulators~\cite{Tsanas2010}. Consequently we derived from MFFC next 3 kinds of perceptual features: LFCC (Linear Frequency Cepstral Coefficients), CMS (Cepstral Mean Subtraction coefficients) and MFCC adjusted to equal loudness curves (as in the case of PLP). In order to provide information complementary to MFCC we tested MSC (Modulation Spectra Coefficients).

Next set of features is based on linear prediction: LPC (Linear Predictive Coefficients), PLP (Perceptual Linear Predictive coefficients), LPCC (Linear Predictive Cepstral Coefficients), LPCT (Linear Predictive Cosine Transform coefficients) and ACW (Adaptive Component Weighted coefficients). In comparison to simple LPC or MFCC the PLP also takes into account adjustment to the equal loudness curve and intensity-loudness power law. The advantage of LPCC and LPCT over ``classic'' LPC is small correlation of values. Finaly the advantage of ACW is that these coefficients are less sensitive to channel distortion.

The last features in this study are ICC (Inferior Colliculus Coefficients) that analyse amplitude modulations in voice using a~biologically-inspired model of the inferior colliculus. All perceptual features were extended by their 1st~order regression coefficients ($\Delta$).

\subsection{Preliminary analysis}

We employed calculation of Spearman's rank sum correlation and mutual information (MI) between feature vectors and resulting speakers' label in order to estimate discrimination power of the vowels separately. Consequently we applied Mann-Whitney~U test and classification based on random forests (RF). Classification results were expressed by ACC, SEN, SPE and trade-off between sensitivity and specificity (TSS) defined as:
\begin{eqnarray}
\mbox{TSS} = 2^{\sin\left(\frac{\pi\cdot\mathrm{SEN}}{2}\right)\sin\left(\frac{\pi\cdot\mathrm{SPE}}{2}\right)}
\end{eqnarray}

Finally to identify perceptual features that can predict values of different clinical tests we calculated Spearman's partial correlations where the effect of patients' age and L-dopa equivalent daily dose was removed.

\subsection{Classification}

In the last step we performed classification with a two-step feature selection. Firstly we reduced set of features to 500 parameters by mRMR (minimum Redundancy Maximum Relevance) and consequently we employed SFFS (Sequential Forward Feature Selection). Three scenarios were considered: individual vowel analysis; classification within each vowel sequence (see Chap.\,\ref{sec:data}); classification using all vowel realizations. In all the cases we used leave-one-out validation.

\section{Experimental results}
\label{sec:results}

\begin{table}
		\scriptsize
		\caption{Individual vowel analysis}
		\label{tab:resind}
		\centering
		\begin{threeparttable}
		\begin{tabular}{l l c c c c c c c}
		\hline
		\hline
		Vowel & Feature & $\rho$ & MI & $p$ & ACC [\%] & SEN [\%] & SPE [\%] & TSS\\
		\hline
		a (s) & 10th ICC (99p) & --0.1356 & 0.0784 & 0.1198 & 71.43 & 72.62 & 69.39 & 1.75\\
		e (s) & 7th LPCC (99p) & --0.0057 & 0.6975 & 0.9498 & 69.92 & 72.62 & 65.31 & 1.71\\
		i (s) & 17th $\Delta$MFCC (1p) & 0.1242 & 0.7349 & 0.1542 & 69.92 & 70.24 & 69.39 & 1.73\\
		o (s) & 10th CMS (std) & 0.0059 & 0.0611 & 0.9477 & \textbf{80.45} & \textbf{78.57} & \textbf{83.67} & \textbf{1.88}\\
		u (s) & 6th ACW (1p) & 0.1003 & 0.7296 & 0.2502 & 72.18 & 75.00 & 67.35 & 1.75\\
		\hline
		a (l) & 9th LFCC (std) & --0.2042 & \textbf{0.7963} & 0.0191 & 75.19 & 75.00 & 75.51 & 1.81\\
		e (l) & 15th MSC (1p) & 0.0938 & 0.0050 & 0.2823 & 66.92 & 65.48 & 69.39 & 1.69\\
		i (l) & 14th ICC (1p) & --0.0503 & 0.0845 & 0.5646 & 68.42 & 69.05 & 67.35 & 1.71\\
		o (l) & 2nd $\Delta$LPCT (median) & 0.0978 & 0.7069 & 0.2619 & 73.68 & 77.38 & 67.35 & 1.76\\
		u (l) & 12th CMS (std) & --0.0842 & 0.0269 & 0.3347 & 77.44 & 73.81 & 83.67 & 1.85\\
		\hline
		a (ll) & 18th $\Delta$LPCC (ir) & 0.0958 & 0.7576 & 0.2720 & 72.18 & 72.62 & 71.43 & 1.76\\
		e (ll) & 11th $\Delta$PLP (99p) & 0.2635 & 0.6129 & 0.0025 & 72.18 & 75.00 & 67.35 & 1.75\\
		i (ll) & 5th PLP (std) & --0.0321 & 0.6460 & 0.7142 & 68.42 & 66.67 & 71.43 & 1.72\\
		o (ll) & 10th $\Delta$CMS (ir) & --0.2038 & 0.7643 & 0.0193 & 75.94 & 73.81 & 79.59 & 1.83\\
		u (ll) & 17th CMS (std) & --0.0486 & 0.0325 & 0.5783 & 72.18 & 76.19 & 65.31 & 1.74\\
		\hline
		a (ls) & 13th ACW (ir) & --0.0093 & 0.7199 & 0.9164 & 73.68 & 79.76 & 63.27 & 1.74\\
		e (ls) & 8th CMS (std) & 0.1887 & 0.0835 & 0.0304 & 72.18 & 64.29 & 85.71 & 1.77\\
		i (ls) & shimmer (local.dB) & \textbf{--0.4064} & 0.7633 & \textbf{0.0000} & 72.18 & 75.00 & 67.35 & 1.75\\
		o (ls) & 3rd ICC (99p) & 0.1324 & 0.0325 & 0.1289 & 69.17 & 71.43 & 65.31 & 1.71\\
		u (ls) & 9th CMS (std) & --0.0191 & 0.0232 & 0.8282 & 75.19 & 69.05 & 85.71 & 1.82\\
		\hline
		\hline
		\end{tabular}
		\begin{tablenotes}
			\scriptsize
      \item[1] $\rho$\,--\,Spearman's rank correlation coefficient; MI\,--\,mutual information; $p$\,--\,significance level (Mann-Whitney~U test; ACC\,--\,classification accuracy; SEN\,--\,sensitivity; SPE\,--\,specificity; TSS\,--\,trade-off between sensitivity and specificity; s\,--\,short vowel pronounced with normal intensity; l\,--\,sustained vowel pronounced with normal intensity; ll\,--\,sustained vowel pronounced with maximum intensity; ls\,--\,sustained vowel pronounced with minimum intensity (not whispering)
    \end{tablenotes}
		\end{threeparttable}
\end{table}

\begin{table}
		\scriptsize
		\caption{Classification results (using feature selection)}
		\label{tab:resall}
		\centering
		\begin{threeparttable}
		\begin{tabular*}{\textwidth}{l @{\extracolsep{\fill}} c c c c c}
		\hline
		\hline
		Vowels & ACC [\%] & SEN [\%] & SPE [\%] & TSS & No.\\
		\hline
		a (s) & 84.21 & 86.90 & 79.59 & 1.90 & 6\\
		e (s) & 81.95 & 82.14 & 81.63 & 1.89 & 8\\
		i (s) & 72.18 & 73.81 & 69.39 & 1.76 & 3\\
		o (s) & 80.45 & 78.57 & 83.67 & 1.88 & 1\\
		u (s) & 85.71 & 86.90 & 83.67 & 1.93 & 6\\
		\hline
		a (l) & 87.22 & 88.10 & 85.71 & 1.94 & 6\\
		e (l) & 75.94 & 78.57 & 71.43 & 1.80 & 7\\
		i (l) & 82.71 & 83.33 & 81.63 & 1.90 & 11\\
		o (l) & 75.19 & 79.76 & 67.35 & 1.77 & 2\\
		u (l) & 77.44 & 73.81 & 83.67 & 1.85 & 1\\
		\hline
		a (ll) & \textbf{91.73} & \textbf{90.48} & \textbf{93.88} & \textbf{1.98} & 8\\
		e (ll) & 78.20 & 83.33 & 69.39 & 1.81 & 3\\
		i (ll) & 78.95 & 82.14 & 73.47 & 1.84 & 6\\
		o (ll) & 81.20 & 80.95 & 81.63 & 1.89 & 3\\
		u (ll) & 72.18 & 76.19 & 65.31 & 1.74 & 1\\
		\hline
		a (ls) & 76.69 & 78.57 & 73.47 & 1.82 & 3\\
		e (ls) & 87.97 & 88.10 & 87.76 & 1.95 & 8\\
		i (ls) & 84.21 & 84.52 & 83.67 & 1.92 & 11\\
		o (ls) & 76.69 & 77.38 & 75.51 & 1.83 & 6\\
		u (ls) & 84.21 & 86.90 & 79.59 & 1.90 & 4\\
		\hline
		all (s) & 80.45 & 78.57 & 83.67 & 1.88 & 1\\
		all (l) & \textbf{91.73} & \textbf{90.48} & \textbf{93.88} & \textbf{1.98} & 9\\
		all (ll) & 81.95 & 79.76 & 85.71 & 1.90 & 7\\
		all (ls) & 90.98 & 91.67 & 89.80 & 1.97 & 11\\
		\hline
		all (s, l, ll, ls) & \textbf{92.48} & \textbf{92.86} & \textbf{91.84} & \textbf{1.98} & 9\\
		\hline
		\hline
		\end{tabular*}
		\begin{tablenotes}
			\scriptsize
      \item[1] ACC\,--\,classification accuracy; SEN\,--\,sensitivity; SPE\,--\,specificity; TSS\,--\,trade-off between sensitivity and specificity; No.\,--\,number of selected features, s\,--\,short vowel pronounced with normal intensity; l\,--\,sustained vowel pronounced with normal intensity; ll\,--\,sustained vowel pronounced with maximum intensity; ls\,--\,sustained vowel pronounced with minimum intensity (not whispering)
    \end{tablenotes}
		\end{threeparttable}
\end{table}

The preliminary results performed by Spearman's rank correlation, mutual information, Mann-Whitney~U test and RF classifier are given in Table~\,\ref{tab:resind}. The results of PD identification based on feature selection can be seen in Table~\,\ref{tab:resall}. Finally the results of Spearman's partial correlations between clinical characteristics and selected features are in Table~\,\ref{tab:parcorr}.

According to the preliminary analysis we can conclude that std of 10th CMS coefficient extracted from short vowel [o] provides the best discrimination power in terms of ACC (80.45\,\%), SEN (78.57\,\%), SPE (83.67\,\%) and TSS (1.88). On the other hand conventional shimmer extracted from sustained vowel [i] pronounced with minimum intensity reached better results of $\rho$ (--0.4064), MI (0.7633) and $p$ (0.0000).

Considering the classification using feature selection, in the first scenario (individual vowel analysis) we can observe the best results in the case of sustained and loudly pronounced vowel [a] (ACC = 91.73\,\%, SEN = 90.48\,\%, SPE = 93.88\,\%, TSS = 1.98). All 8 selected features were perceptual. In the case of second scenario (classification within each vowel sequence) the best results provided sustained vowels pronounced with natural intensity (ACC = 91.73\,\%, SEN = 90.48\,\%, SPE = 93.88\,\%, TSS = 1.98), where all 9 selected features were perceptual as well. It was proved that in order to get best classification results (ACC = 92.48\,\%, SEN = 92.86\,\%, SPE = 91.84\,\%, TSS = 1.98) it is advantageous to use all 4 sets of vowels.

In our recent study we found out that sustained vowels pronounced with minimum intensity can be good speech tasks for detection of improper vocal folds vibration (measured by features based on empirical mode decomposition)\cite{Smekal2015}. In the case of perceptual analysis we observe that loudly pronounced features are better candidates to analyse. We explain this by substantiality of perception. Theoretically longer and more intense stimuli results in better perception.


Finally we have proved that perceptual features significantly correlate ($p < 0.0001$) with different clinical information like UPDRS III (Unified Parkinson's disease rating scale, part III: Motor Examination), UPDRS IV (part IV: Complications of Therapy), RBDSQ (The REM sleep behavior disorder screening questionnaire), FOG (Freezing of gait questionnaire), NMSS (Non-motor symptoms scale), BDI (Beck depression inventory) and MMSE (Mini-mental state examination). This means that they can be used for estimation of these scores.

\begin{table}
		\scriptsize
		\caption{Spearman's partial correlations between clinical characteristics and selected features (after removal of age and LED effect)}
		\label{tab:parcorr}
		\centering
		\begin{tabular}{l l c c}
		\hline
		\hline
Clinical info & Feature & $\rho$ & $p$\\
\hline
PD duration & i (l): 15th CMS (std) & --0.4369~~ & $3.25\cdot10^{-5}$\\
UPDRS III & i (l): 1st $\Delta$PLP (1p) & --0.5174 & $6.98\cdot10^{-7}$\\
UPDRS IV & e (ll): 5th $\Delta$MFCC (ir) & 0.4572 & $1.23\cdot10^{-5}$\\
RBDSQ & u (ls): 13th $\Delta$MFCC (99p) & 0.4906 & $2.16\cdot10^{-6}$\\
FOG & a (ls): 6th MFCC (std) & --0.4476 & $1.96\cdot10^{-5}$\\
NMSS & a (ll): 12th LPC (99p) & 0.4616 & $1.25\cdot10^{-5}$\\
BDI & u (s): 3rd $\Delta$LPCT (1p) & 0.5832 & $1.25\cdot10^{-6}$\\
MMSE & i (l): 20th MFCC (99p) & --0.4719 & $5.55\cdot10^{-5}$\\
		\hline
		\hline
		\end{tabular}
\end{table}

\section{Conclusion}
\label{sec:conclusion}

In this paper we perceptualy analysed phonation of 84 PD patients and 49 gender and age matched controls. We achieved all goals of this work: 1) We have proved that perceptual features outperform the conventional ones in terms of discrimination power. 2) From a wide range of perceptual features we have found out that those based on CMS (derived from MFCC) better quantify the signs of hypokinetic dysarthria. 3) We have shown that it is advantageous to perform perceptual analysis of loud sustained vowels. 4) In the case of each considered clinical score we identified a perceptual feature that can be used for its estimation.

In the near future we would like to move further, perceptualy analyse another speech tasks (spontaneous speech, read sentences, etc.) and focus on each gender individually.

\section*{Acknowledgment}
Research described in this paper was financed by the National Sustainability Program under grant LO1401 and by projects NT13499 (Speech, its impairment and cognitive performance in Parkinson's disease), COST IC1206, project ``CEITEC, Central European Institute of Technology'': (CZ.1.05/1.1.00/02.0068), FEDER and Ministerio de Econom\'{i}a y Competitividad TEC2012-38630-C04-03.

%
%
\bibliographystyle{splncs03}
\bibliography{NOLISP2015}

\begin{thebibliography}{10}
\providecommand{\url}[1]{\texttt{#1}}
\providecommand{\urlprefix}{URL }

\bibitem{Arnold2014}
Arnold, C., Gehrig, J., Gispert, S., Seifried, C., Kell, C.A.: Pathomechanisms
  and compensatory efforts related to {Parkinsonian} speech. Neuroimage Clin
  4(0),  82--97 (2014)

\bibitem{Bocklet2011}
Bocklet, T., Noth, E., Stemmer, G., Ruzickova, H., Rusz, J.: Detection of
  persons with {Parkinson's} disease by acoustic, vocal, and prosodic analysis.
  In: Automatic Speech Recognition and Understanding (ASRU), 2011 IEEE Workshop
  on. pp. 478--483 (2011)

\bibitem{Brodal2003}
Brodal, P.: The Central Nervous System: Structure and Function. Oxford
  University Press, 3 edn. (2003)

\bibitem{Chenausky2011}
Chenausky, K., MacAuslan, J., Goldhor, R.: Acoustic analysis of pd speech.
  Parkinson's Disease  2011,  1--13 (2011)

\bibitem{Eliasova2013}
Eliasova, I., Mekyska, J., Kostalova, M., Marecek, R., Smekal, Z., Rektorova,
  I.: Acoustic evaluation of short-term effects of repetitive transcranial
  magnetic stimulation on motor aspects of speech in {Parkinson's} disease. J
  Neural Transm  120(4),  597--605 (2013)

\bibitem{Mekyska2011b}
Mekyska, J., Smekal, Z., Kostalova, M., Mrackova, M., Skutilova, S., Rektorova,
  I.: Motor aspects of speech imparment in {Parkinson's} disease and their
  assessment. Cesk Slov Neurol N  74(6),  662--668 (2011)

\bibitem{Orozco2013b}
Orozco-Arroyave, J., Arias-Londono, J., Vargas-Bonilla, J., Noth, E.:
  Perceptual analysis of speech signals from people with {Parkinson's} disease.
  In: Natural and Artificial Models in Computation and Biology, Lecture Notes
  in Computer Science, vol. 7930, pp. 201--211. Springer Berlin Heidelberg
  (2013)

\bibitem{Rusz2011d}
Rusz, J., Cmejla, R., Ruzickova, H., Ruzicka, E.: Quantitative acoustic
  measurements for characterization of speech and voice disorders in early
  untreated {Parkinson's} disease. J Acoust Soc Am  129(1),  350--367 (2011)

\bibitem{Sapir2008}
Sapir, S., Ramig, L., Fox, C.: Speech and swallowing disorders in {Parkinson}
  disease. Curr Opin Otolaryngol Head Neck Surg  16(3),  205--210 (2008)

\bibitem{Skodda2013}
Skodda, S., Grunheit, W., Mancinelli, N., Schlegel, U.: Progression of voice
  and speech impairment in the course of {Parkinson's} disease: A longitudinal
  study. Parkinson's Disease  2013,  1--8 (2013)

\bibitem{Skodda2010}
Skodda, S., Visser, W., Schlegel, U.: Short- and long-term dopaminergic effects
  on dysarthria in early {Parkinson's} disease. J Neural Transm  117,  197--205
  (2010)

\bibitem{Smekal2015}
Smekal, Z., Mekyska, J., Galaz, Z., Mzourek, Z., Rektorova, I., Faundez-Zanuy,
  M.: Analysis of phonation in patients with {Parkinson's} disease using
  empirical mode decomposition. In: 2015 International Symposium on Signals,
  Circuits and Systems (ISSCS). pp. 1--4 (2015)

\bibitem{Tsanas2010}
Tsanas, A., Little, M.A., McSharry, P.E., Ramig, L.O.: Nonlinear speech
  analysis algorithms mapped to a standard metric achieve clinically useful
  quantification of average {Parkinson's} disease symptom severity. J R Soc
  Interface  8(59),  842--855 (2010)

\end{thebibliography}

\end{document}